\documentclass[prb,twocolumn,showpacs,preprintnumbers,amsssymb]{revtex4}
\usepackage{graphicx}
\usepackage{dcolumn}
\usepackage{bm}

\begin{document}

\title{Quasiparticle dynamics and in-plane anisotropy in YBa$_{2}$Cu$_{3}$O$%
_{y}$ system near onset of superconductivity}
\author{Y.-S. Lee,$^{1}$ Kouji Segawa,$^{2}$ Yoichi Ando,$^{2}$ and D. N.
Basov$^{1}$}
\affiliation{$^{1}$Department of Physics, University of California at San Diego, La
Jolla, California 92093-0319\\
$^{2}$Central Research Institute of Electric Power Industry, Komae, Tokyo
201-8511, Japan}
\date{\today }

\begin{abstract}
We report on an infrared study of carrier dynamics within the CuO$_{2}$
planes in heavily underdoped detwinned single crystals of YBa$_{2}$Cu$_{3}$O$%
_{y}$. In an effort to reveal the electronic structure near the onset of
superconductivity, we investigate the strong anisotropy of the
electromagnetic response due to an enhancement of the scattering rate along
the $a$-axis. We propose that the origin of this anisotropy is related to a
modulation of the electron density within the CuO$_{2}$ planes.
\end{abstract}

\maketitle

The physics of Mott-Hubbard (MH) insulators is at the focus of current
research in part because of a variety of enigmatic ground states that can be
initiated through doping of these systems.\cite{MIT98} A particularly
interesting example is high-$T_{c}$ superconductivity which is triggered by
doping of MH insulating cuprates and is believed to originate from strong
correlations among the doped holes.\cite{Orenstein00} To elucidate the
unconventional superconductivity, it is therefore imperative to understand
the nature of the conducting state derived from the Mott insulator. Thus
motivated, we have carried out systematic studies of the electromagnetic
response of a prototypical high-$T_{c}$ superconductor YBa$_{2}$Cu$_{3}$O$%
_{y}$ (YBCO) over a broad region of the phase diagram across the
antiferromagnetic (AF) transition and onset of superconductivity.

One of the intriguing properties of heavily underdoped YBCO is the in-plane
anisotropy in the DC resistivity.\cite{Ando02} This effect is likely to be
distinct from the anisotropy seen in the YBCO system near optimal doping.%
\cite{Ando02,Friedmann90,Harris01} The latter is attributed to the
one-dimensional (1-D) Cu-O chains along the $b$-axis which are capable of
contributing both to the DC transport and to the superfluid density, \cite%
{Basov95,wachter94} provided the disorder in the chains is weak. Although
the direct role of chains in the anisotropy of weakly doped phases has not
been explicitly ruled out, this scenario is rather remote because
considerable defects in the chains are inevitable in oxygen deficient
crystals. In searching for alternative mechanisms of the anisotropy, it is
prudent to take into consideration anisotropic transport and infrared
response within the CuO$_{2}$ plane in a nearly tetragonal chain-free
cuprate La$_{2-x}$Sr$_{x}$CuO$_{4}$ (LSCO).\cite{Ando02,Dumm03} It has been
argued that the directional dependence of quasiparticle dynamics in LSCO may
be related to the intrinsic tendency of doped MH insulators towards spin
and/or charge self-organization in real space. Similar spin/charge
modulation has also been observed in YBCO from neutron and x-ray scattering
experiments.\cite{Dai98,Mook99,Mook00,Sinha02} Infrared (IR) spectroscopy is
ideally suited for elaborating in-plane anisotropy in the heavily underdoped
YBCO because reflectance measurements with polarized light are capable of
separating the contribution of the 1-D chains from the response of the CuO$%
_{2}$ planes.
\begin{figure}[b]
\includegraphics[width=0.45\textwidth]{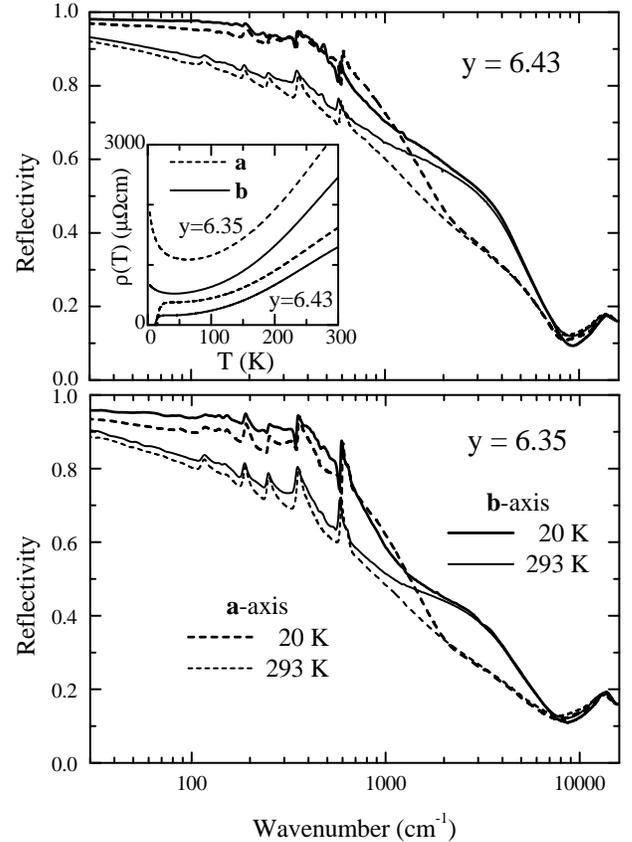}
\caption{$T$-dependent $R(\protect\omega )$ for (a) $y=6.43$ and (b) $y=6.35$%
. Inset in (a) shows DC resistivity curves for $y$ = 6.43 and 6.35 crystals.}
\end{figure}

Here we report on the study of the optical spectra for the heavily
underdoped YBCO with oxygen contents $y$ = 6.30, 6.35, 6.40, and 6.43, grown
by a conventional flux method and detwinned under uniaxial pressure.\cite%
{Segawa01} The $y$ = 6.43 and 6.40 samples are superconducting with
transition at $T_{c}$ $\sim $ 13 K and $\sim $ 2 K, respectively, determined
from the DC resistivity measurement shown in the inset of Fig. 1(a).\cite%
{Segawa01} At lower dopings the ground state of YBCO is antiferromagnetic.%
\cite{Jorgensen90,Lavrov99} Annealing under the uniaxial pressure aligns
chain segments along the $b$-axis in non-superconducting YBCO crystals. All
samples show an in-plane anisotropy in resistivity, with higher conductivity
along the $b$-axis.\cite{Ando02} Reflectivity spectra $R(\omega )$ at near
normal incidence were measured with polarized light at frequencies from 20
to 48000 cm$^{-1}$ and at temperatures from 20 to 293 K. According to the
Hagen-Rubens relation, $R(\omega )\sim \sqrt{\omega \rho _{\text{DC}}}$,
where $\rho _{\text{DC}}$ is the DC resistivity value. As shown in Fig. 1,
the far-IR reflectivities along the $b$-axis are higher than those along the 
$a$-axis, which is consistent with the anisotropy in the DC resistivity. In
addition, the mid-IR $R(\omega )$ is enhanced along the $b$-axis. This
behavior is associated with the optical response of chain fragments aligned
along the $b$-axis, which will be described below. With $T$ decreasing, the
low frequency $R(\omega )$ increase considerably whereas the change in the
mid-IR is negligible. The complex optical conductivity spectra, $\widetilde{%
\sigma }(\omega )=\sigma _{1}(\omega )+i\sigma _{2}(\omega )$, were obtained
from the measured $R(\omega )$, using Kramers-Kronig (KK) transformation.
The KK-derived results are consistent with those obtained independently by
spectroscopic ellipsometry in the near-IR and visible region.
\begin{figure}[tbp]
\includegraphics[width=0.45\textwidth]{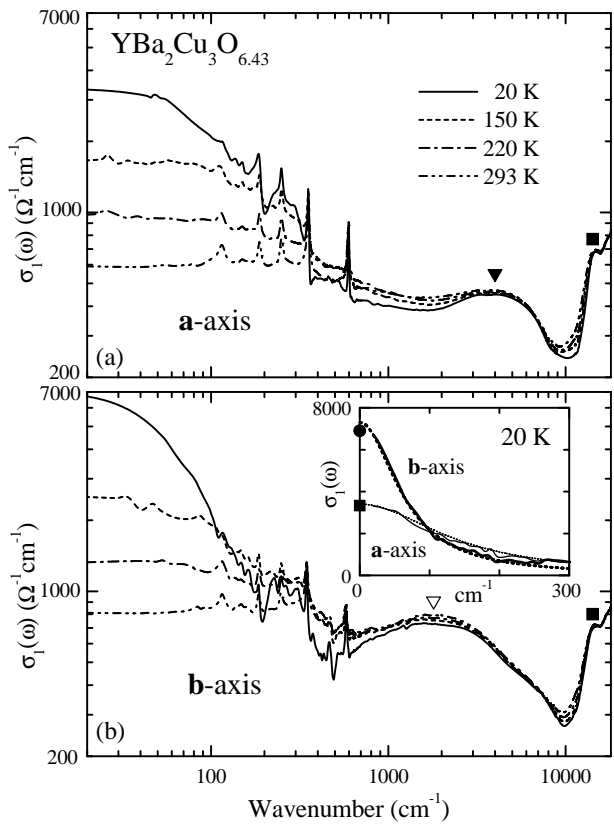}
\caption{$T$-dependent $\protect\sigma _{1}(\protect\omega )$ of the $y=6.43$
YBCO along (a) the $a$-axis and (b) the $b$-axis. The circle and the square
symbols represent the positions of MIR resonances and CT excitations,
respectively. In the inset in (b), the $\protect\sigma _{1}(\protect\omega )$
with the phonon subtracted are used for the Drude fitting (the dotted
lines). Note that the fits are adequate for the frequency range exceeding $%
\Gamma $ parameters. The symbols represent the DC resistivity values.}
\end{figure}
\begin{figure}[tbp]
\includegraphics[width=0.45\textwidth]{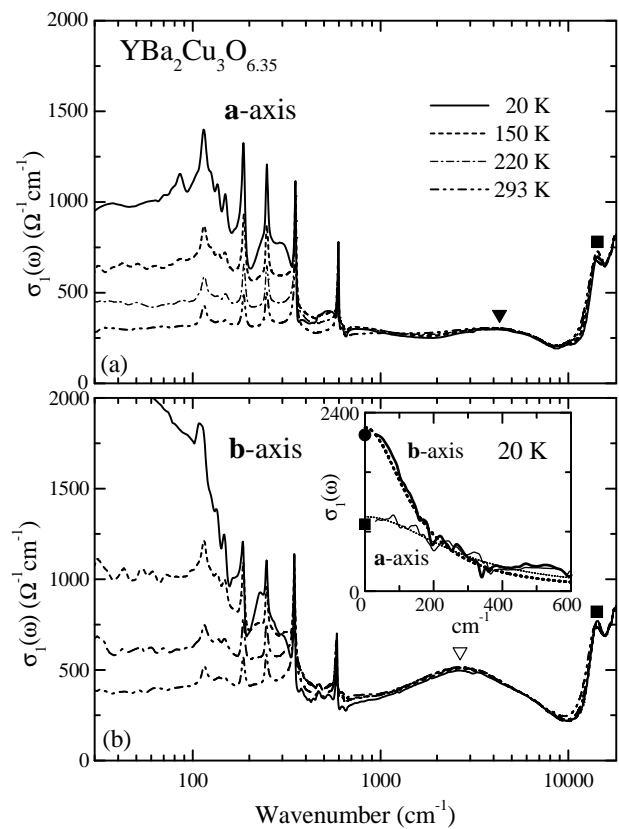}
\caption{$T$-dependent $\protect\sigma _{1}(\protect\omega )$ of the $y=6.35$
YBCO in (a) the $a$-axis and (b) the $b$-aixs. The circle and the square
symbols represent the positions of the MIR resonances and the CT
excitations, respectively. In the inset (b), the $\protect\sigma _{1}(%
\protect\omega )$ with the phonon subtracted are used for the Drude fitting
(the dotted lines). The symbols represent the DC resistivity values.}
\end{figure}

Figures 2(a) and 2(b) show the dissipative part of the optical conductivity $%
\sigma _{1}(\omega )$ for the $y$ = 6.43 sample along the $a$- ($\sigma
_{1,a}(\omega ))$ and the $b$-axis ($\sigma _{1,b}(\omega ))$, respectively.
The optical spectra in both polarizations show qualitatively common
features, including a charge transfer (CT) excitation seen around 14,000 cm$%
^{-1}$ and substantial electronic spectral weight within the CT gap. The
intra-gap spectral weight is comprised of two separate absorption features:
a coherent mode at lowest frequencies ($\omega <600$ cm$^{-1})$ followed by
a mid-IR resonance. The distinct character of the two absorption features is
most evident in the low-$T$ data: the mid-IR structure is virtually
independent of $T$, whereas the coherent mode significantly narrows at low-$T
$ leading to a minimum in $\sigma _{1}(\omega )$ between the two components.
This two-component optical conductivity is also observed in the other YBCO
samples (Fig. 3) and is a general characteristic of weakly doped cuprates.%
\cite{book1} Apart from this electronic contribution, the conductivity
spectra also reveal sharp peaks due to IR-active phonons. The phonon
frequencies show noticeable differences in the $a$- and $b$-axis data
confirming that the studied samples are detwinned. 

Coherent low-$\omega $ features in the optical response of all YBCO crystals
are well described with the AC Drude formula, $\widetilde{\sigma }(\omega
)=(\omega _{p}^{2}/4\pi )/(\Gamma -i\omega )$, where $\omega _{p}$ and $%
\Gamma $ are the plasma frequency and the scattering rate of carriers,
respectively. The fits are presented with the dotted lines in the insets of
Figs. 2(b) and 3(b) with the fitting parameters reported in Table I.
Interestingly, the Drude description holds for the $y=6.35$ crystal at 20 K,
which is less than the N\'{e}el temperature for this particular composition.%
\cite{comment3} The coherent response of the $y=6.30$ sample at the lowest $%
T $ is modified compared to the simple Drude formula: the electronic
conductivity reveals a weak peak near 100 cm$^{-1}$ suggestive of the
carrier localization.

An outstanding property of heavily underdoped YBCO is anisotropy in the
coherent modes with a marked increase of the conductivity along the $b$%
-direction. A quick inspection of the Drude parameters in Table I reveals
that this effect is primarily produced by the anisotropy of $\Gamma $
whereas the anisotropy of $\omega _{p}$ is fairly small. An independent
confirmation of this finding is provided by an examination of the electronic
spectral weight $N_{\text{eff}}(\omega )=\int_{0}^{\omega }\sigma
_{1}(\omega ^{\prime })d\omega ^{\prime }$. The oscillator strength sum rule
suggests that integration over the frequency region of the coherent mode ($%
\omega \sim 600$ cm$^{-1}$) provides an accurate estimate of $\omega _{p}^{2}
$. The spectra presented in Figs. 4(a) and 4(b) show that the anisotropy of
the plasma frequency $\omega _{p,b}/\omega _{p,a}=\sqrt{N_{\text{eff,}b}/N_{%
\text{eff,}a}}$ does not exceed 1.12 for the $y=6.35$ sample and 1.07 for
the $y$ = 6.43 specimen at $\omega =600$ cm$^{-1}$. These results are quite
comparable with the Drude fitting results in Table I. As $\omega \rightarrow
0$, the $N_{\text{eff,}b}(\omega )/N_{\text{eff,}a}(\omega )$ values
approach the ratio of the DC conductivities $\sigma _{\text{DC,}b}/\sigma _{%
\text{DC,}a}$, indicating that the optical results are also consistent with
the transport data. The dominant role of $\Gamma $ in the anisotropic
carrier dynamics of YBCO is further confirmed by the data in the inset of
Fig. 4(b) where we display the $T$-dependent $\Gamma _{a}/\Gamma _{b}$
extracted from Drude fits along with the DC resistivity ratios $\rho _{\text{%
DC,}a}/\rho _{\text{DC,}b}$. Close agreement between the two data sets
establishes that the significant anisotropy in $\Gamma $ is primarily
responsible for the resistivity anisotropy within the CuO$_{2}$ plane in
weakly doped YBCO compounds. 
\begin{table}[tbp]
\caption{Summaries of the Drude fitting parameters, the plasma frequency $%
\protect\omega _{p}$ and the scattering rate $\Gamma $, for $\protect\sigma %
_{1}(\protect\omega )$ of YBa$_{2}$Cu$_{3}$O$y$ at 20 K along the $a$- and $b
$- axis. For $y=6.30$, we used the data at 80 K just above the onset $T$ of
an insulating-like state. Using the measured Hall coefficient at 300K, the
mean free path $\ell $ at the corresponding $T$ are also estimated with a
three dimensional free electron model.}%
\begin{ruledtabular}
\begin{tabular}[t]{cc|cccccc}
\multicolumn{2}{c}{$y$} & \multicolumn{2}{c}{$\omega _p$ (cm$^{-1}$)} & 
\multicolumn{2}{c}{$\Gamma $ (cm$^{-1}$)} & \multicolumn{2}{c}{$\ell $ (\AA )
} \\ 
\multicolumn{2}{c}{} & $a$ & $b$ & $a$ & $b$ & $a$ & $b$ \\ 
\multicolumn{2}{c}{6.43} & 5250 & 5340 & 135 & 65 & 57 & 121 \\ 
\multicolumn{2}{c}{6.40} & 4350 & 4420 & 210 & 105 & 31 & 64 \\ 
\multicolumn{2}{c}{6.35} & 4100 & 4300 & 280 & 140 & 24 & 52 \\ 
\multicolumn{2}{c}{6.30} & 3630 & 3890 & 430 & 270 & 14 & 25
\end{tabular}
\end{ruledtabular}
\end{table}

What is the origin of the anisotropic scattering? Searching for plausible
causes of this enigmatic effect, it is imperative to access the possible
role of Cu-O chains as well as that of the orthorhombicity of YBCO. As
pointed out above, highly ordered chains in the (nearly) stoichiometric YBa$%
_{2}$Cu$_{3}$O$_{6.95}$ and YBa$_{2}$Cu$_{4}$O$_{8}$ crystals contribute
directly to DC transport.\cite{wachter94} This contribution is reflected in
strong enhancement of the $b$-axis plasma frequency and in the anisotropy of 
$\sigma _{\text{DC,}b}/\sigma _{\text{DC,}a}$ that appears to scale with $%
\omega _{p,b}/\omega _{p,a}$.\cite{Basov95,wachter94} Notably, YBCO crystals
with $T_{c}\simeq 90$ K containing significant disorder on the chain sites
(due to non-optimized growth and/or annealing conditions) \textit{do not}
reveal substantial $\sigma _{DC}$ anisotropy.\cite{disorder1} In this case,
the $b$-axis conductivity shows a strong resonance at $\omega \simeq
1500-2000$ cm$^{-1}$, usually assigned to carrier localization in the Cu-O
chain segments.\cite{Cooper93,Rotter91,Takenaka92} Examining the mid-IR
response of the weakly doped YBCO in light of these earlier results, we
first note substantial anisotropy of the mid-IR conductivity which can be
readily traced in $R(\omega )$. The mid-IR resonances in the $a$-axis
spectra are located at $\sim $ 4200 cm$^{-1}$ without any significant $y$%
-dependence in the heavily underdoped region. This feature reflects the
intrinsic electronic structure of the doped CuO$_{2}$ planes and is rather
insensitive to a particular host material.\cite{Dumm03,book1,Uchida91} On
the other hand, the mid-IR resonances in the $b$-axis are located at much
lower frequencies: $\sim $ 1700 cm$^{-1}$ at $y=6.43$ and $\sim $ 3100 cm$%
^{-1}$ at $y=6.30$. The spectral weight of these $b$-axis features
significantly exceeds that of the resonances observed in the $a$-axis
spectra. To obtain further insight into the chain response, we checked $%
\sigma _{1,b}(\omega )-\sigma _{1,a}(\omega )$ which are attributable to the
chains. These differential spectra reveal an over-damped oscillator at 1500
- 2600 cm$^{-1}$ and are comparable to the disordered chain response in the
YBCO system at higher doping.\cite{Cooper93,Rotter91,Takenaka92} As $y$
decreases and the density of oxygen defects in the chains is enhanced, the
peaks shift to higher frequency while their spectral weight is reduced.
These observations are in accord with the theoretical analysis of the
localized state of the oxygen-deficient chains in YBCO system.\cite{Franco03}
The notion of carrier localization in the Cu-O chain segments is consistent
with the experimental fact that no significant anisotropy in $\omega _{p}$
is found in our data. We therefore conclude that the chains in the studied
specimen are unable to maintain conductivity in the DC or low-$\omega $
limit. 
\begin{figure}[tbp]
\includegraphics[width=0.48\textwidth]{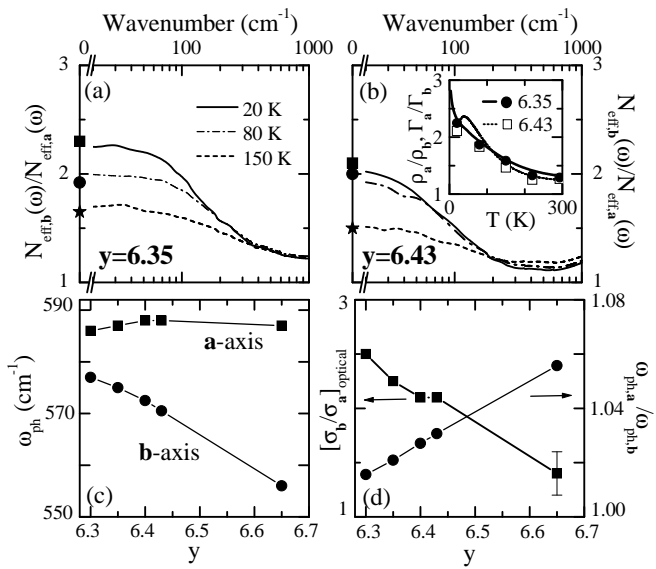}
\caption{Top panels: $T$-dependent $N_{\text{eff,}b}(\protect\omega )/N_{%
\text{eff,}a}(\protect\omega )$ at (a) $y=6.30$ and (b) $y=6.43$. The
square, the circle, and the star symbols correspond to the measured $\protect%
\sigma _{\text{DC,}b}/\protect\sigma _{\text{DC,}a}$ values at 20 K, 80K,
and 150 K, respectively. Low panels display the evolution of the anisotropy
of the phonon modes. (c) At 293 K, the frequencies of phonons positioned at
the highest frequency at $y=$ 6.30, 6.35, 6.40. 6.43, and 6.65. (d) $y$%
-dependences of $\protect\omega _{\text{ph,}a}/\protect\omega _{\text{ph,}b}$
and [$\protect\sigma _{b}/\protect\sigma _{a}]_{\text{optical}}\equiv \lim_{%
\protect\omega \rightarrow 0}\protect\sigma _{1,b}/\protect\sigma _{1,a}$.
For $y=6.65$, we choose the value at 65 K just above $T_{c}$. In the inset
of (b), the dotted and the solid lines represent the $\protect\rho _{a}/%
\protect\rho _{b}$ curves at $y=6.43$ and $y=6.35$, respectively. The square
($y=6.43$) and the circle ($y=6.35$) symbols correspond to the $T$-dependent 
$\Gamma _{a}/\Gamma _{b}$ values multiplied by 1.05 and 1.1, respectively.}
\end{figure}

We now turn to the possible impact of orthorhombicity on anisotropic
quasiparticles dynamics. As pointed out above, the bulk orthorhombicity of
the studied specimens is supported by the anisotropy of the phonon
frequencies. Their doping trends are shown in Fig. 4(c) and 4(d), where we
display the $y$-dependence of the stretching mode frequencies $\omega _{%
\text{ph}}$, which are most sensitive to the variation of the lattice
constants.\cite{Tajima91} Comparing the $\omega _{\text{ph,}a}/\omega _{%
\text{ph,}b}$ ratios with the anisotropy of the DC conductivity (or AC
conductivity in the $\omega \rightarrow 0$ limit), one notices that the
latter increases as the YBCO system progresses towards the tetragonal phase (%
$\omega _{\text{ph,}a}/\omega _{\text{ph,}b}\rightarrow 1$).\cite%
{Jorgensen90} Interestingly, the higher $\omega _{\text{ph,}a}$ values
indicate that the lattice constant in the $a$-axis is shorter than that in
the $b$-axis. According to the tight binding picture, one would expect to
find an enhancement of the electronic spectral weight (proportional to
hopping integrals) in the polarization of $\mathbf{E}$ vector along the
shorter axis, which is not realized in the $N_{\text{eff}}$ analysis
discussed above. We therefore conclude that structural orthorhombicity can
be safely excluded from factors defining anisotropic electronic response of
weakly doped YBCO. 

What are the microscopics behind the anisotropic scattering within the CuO$%
_{2}$ plane leading to an enhancement of the conductivity along the
direction of chain segments? It is rather unlikely that the chains directly
contribute to the anisotropic in-plane $\Gamma $ because scattering by the
out-of-plane defects involves only a small momentum transfer which is of
little consequence for transport phenomena.\cite{Kee01} With this
possibility being eliminated it is appealing to explore indirect impact of
chains through an \textquotedblleft imprint\textquotedblright\ of the charge
density in the CuO$_{2}$ planes recently observed through both x-ray
experiments\cite{Sinha02} and nuclear resonance.\cite{grevin00}. These
latter studies indicate that chain segments act as a template to induce
(stripe-like) charge modulation in the neighboring planes. This modulated
state may act as a scattering source enhancing the $\Gamma $ for carriers
propagating along the modulation direction ($a$-axis) in accord with our
observations. We note that the ordered chain segments form \textquotedblleft
patches\textquotedblright\ elongated along the $b$-axis with the aspect
ratio $\sim $ 3:1,\cite{greven00} which are comparable to the anisotropy of
the mean free path within the CuO$_{2}$ plane reported in Table I. The
charge modulation may be related to the tendency of doped MH insulators to
form self-organized stripe-like phases when both spin and charge ordering
are expected. Indeed the $T$-dependence of the anisotropy of $\Gamma $
appears to be in accord with that of the strength of the stripe-related 
\textit{magnetic} structure seen in neutron scattering measurements.\cite%
{Dai98,Mook99}

The experiments reported here provide important insights into the nature of
the modulated electronic state. First, the charge modulation appears to be
only loosely coupled to the lattice because no phonon zone-folding effects
can be readily identified within the signal-to-nose ratio in our data.\cite%
{bernhard} Second, the amplitude of the charge modulations must be rather
weak because the electronic transport within the CuO$_{2}$ planes retains
its two-dimensional character even at the lowest $T$ where the anisotropy is
maximized. This anisotropic transport in the metallic state has to be
contrasted with the case of the weakly doped LSCO system. In the latter
materials, the anisotropic conductivity is accompanied with the development
of localization peak in $\sigma _{1}(\omega )$ spectra that is in
qualitative agreement with the notion of 1-D confinement of carrier motion.%
\cite{Dumm03} A common aspect of both cuprates is that the off-plane
structural distortions (chain fragments in YBCO and octahedra tilts in LSCO)
appear to play a preeminent role in stabilizing charge non-uniformity within
the CuO$_{2}$ planes. Charge modulations triggered by these two stabilizing
methods are different: enhanced electron density is colinear with respect to
Cu-O bonds in the case of YBCO\ and diagonal in LSCO. Nevertheless, in both
cases we observe an enhancement of the far-IR conductivity across the
modulation direction signaling that this effect is the intrinsic property of
weakly doped CuO$_{2}$ planes. We remark that in YBCO systems the
(stripe-like) charge modulations are found here to coexist with
superconductivity in contrast to the experimental situation in LSCO.\cite%
{ichikawa00} This observation is significant because the stripes framework
offers an interesting prospective on the pairing mechanism in high-$T_{c}$
cuprates.\cite{emery97}

In conclusion, our infrared studies of heavily underdoped YBCO reveal that
the in-plane optical spectra below the charge transfer gap are comprised of
the well-separated Drude-like coherent mode and mid-IR resonance. We observe
substantial anisotropy of the electronic transport within the CuO$_{2}$
planes due to the enhanced scattering rate along the $a$-axis. The chain
segments and their ordering appear to play a profound role in the
anisotropic transport by imprinting charge modulations in the CuO$_{2}$
planes.

We wish to thank S. A. Kivelson for his fruitful discussions and valuable
comments. This research was supported by the US Department of Energy Grant
No. DE-FG0300ER45799. Y.-S.L. was partially supported by the Post-doctoral
Fellowship Program of Korea Science Engineering Foundation (KOSEF).

\newpage


\begin{thebibliography}{99}
\bibitem{comment3} The low frequency localization mode in the FIR region,
which is clearly observed at $y=6.30$, is expected to occur far below 20 K,
consistent with the insulating-like behavior in $\rho (T)$.
\end{thebibliography}
\end{document}